\documentclass[a4paper,11pt]{article}
\usepackage{pos}
\usepackage{subcaption}
\usepackage{booktabs}

\title{Isospin-$\frac{1}{2}$ $D\pi$ scattering and the $D_0^*$ resonance}
\author*[a]{Haobo Yan}
\author[a,b,c]{Chuan Liu}
\author[d,e]{Liuming Liu}
\author[f]{Yu Meng}
\author[d,e]{Hanyang Xing}

\affiliation[a]{School of Physics, Peking University, Beijing 100871, China}
\affiliation[b]{Center for High Energy Physics, Peking University, Beijing 100871, China}
\affiliation[c]{Collaborative Innovation Center of Quantum Matter, Beijing 100871, China}
\affiliation[d]{Institute of Modern Physics, Chinese Academy of Sciences, Lanzhou 730000, China}
\affiliation[e]{University of Chinese Academy of Sciences, Beijing 100049, China}
\affiliation[f]{School of Physics and Microelectronics, Zhengzhou University, Zhengzhou, Henan 450001, China}

\emailAdd{haobo@stu.pku.edu.cn}

\abstract{
Preliminary lattice QCD results for $D\pi$ scattering in isospin $I=\frac{1}{2}$ channel are presented. Utilizing the $N_f=2+1$ Wilson-Clover configuration at two volumes ($L^3 \times T=32^3 \times 96$ and $48^3 \times 96$) with the same lattice spacing ($a=0.07746(18)$ fm) and pion mass ($m_\pi \approx 303$ MeV), various two-particle operators in both the COM and the moving frames are constructed and the corresponding finite-volume spectra are determined from their correlation functions. The $S$ and $P$-wave scattering phase shifts are then extracted using the L\"uscher approach, assuming negligible contributions from higher partial waves. A virtual state associated with the $D_0^*$ is also identified.
}

\FullConference{The 40th International Symposium on Lattice Field Theory (Lattice 2023)\\
July 31st - August 4th, 2023\\
Fermi National Accelerator Laboratory\\}

\begin{document}
\maketitle

\section{Introduction}

To shed more light on the understanding of the charm-light meson sector, efforts have been devoted to the study of the scalar resonance $D_0^*$, which is discovered and verified by many experiments~\cite{Abe2004, Link2004, Aubert2009, Aaij2015, Aaij2015a}. Still, various puzzles remain. For example, the scalar resonance $D_0^* (2300)$ lies so close to its strange partner $D_{s0}^* (2317)$, which is attributed within the quark model to the strong coupling to the threshold~\cite{Chen2017}. There are also suggestions of a two-pole structure in this channel from unitarized chiral perturbation theory~\cite{Albaladejo2017}, however, this structure needs to be tested at different quark masses. It is therefore crucial to study this scenario more systematically within the non-perturbative framework such as lattice QCD.

Lattice QCD computations on hadron-hadron scattering rely on the
so-called L\"uscher method and its extensions, within this formalism one relates 
the finite-volume spectra obtained from lattice QCD to the infinite volume scattering amplitudes in the continuum. For the two particle systems, this is by now rather well established for both elastic and coupled channels, see e.g. the review~\cite{Briceno2018} and references therein.

Previous lattice works on $D\pi$ scattering were conducted on an $N_{\text{f}}=2$, $m_{\pi} \approx 266$ MeV~\cite{Mohler2013} lattice. Afterwards, the Hadron Spectrum Collaboration performed a coupled-channel analysis using two sets of anisotropic lattice ensembles with many operators~\cite{Moir2016,Gayer2021}. In this work, we report preliminary lattice results on $D\pi$ scattering using a new set of symmetric $N_{\text{f}}=2+1$ Wilson-clover lattice ensembles generated by CLQCD Collaboration at an intermediate $m_{\pi} \approx 303$ MeV to further examine the behavior of the $D_0^*$ pole as a function of $m_{\pi}$.

\section{Operator construction}
To constrain the phase shifts more effectively near the threshold, both single and two-hadron operators need to be constructed to interpolate the full spectra in a particular channel. It is also crucial to construct operators in both the COM frame and as many as possible moving frames to have more points populating the relevant kinematic region. For the construction of the single-hadron operators, the helicity strategy proposed by the Hadron Spectrum Collaboration~\cite{Dudek2010, Thomas2012} is applied. The construction is divided into $3$ steps: First, the operators with specific angular momentum in the infinite volume are constructed by conventional Clebsch-Gordan (CG) coefficients,
\begin{equation}
\mathcal{O}^{J M}(\vec{P}) \sim \sum_{m_1, m_2, m_3, \ldots} \operatorname{CGs}\left(m_1, m_2, m_3, \ldots\right) \sum_{\vec{x}} e^{i \vec{P} \cdot \vec{x}} \bar{\psi}(\vec{x}, t) \Gamma_{m_1} \overleftrightarrow{D}_{m_2} \overleftrightarrow{D}_{m_3} \psi(\vec{x}, t),
\end{equation}
where $\vec{P}$ is the center-of-mass frame momentum and the matrices $\Gamma_m$ and covariant derivatives $\overleftrightarrow{D}_m$ carry irreducible spherical vector index $m=0,\pm$. Second, helicity operators are constructed by performing an $SU(2)$ rotation $R$ that takes the three-vector $(0,0,|\vec{P}|)$ to the general chosen $\vec{P}$,
\begin{equation}
\mathbb{O}^{J P \lambda}(\vec{P})=\sum_M \mathcal{D}_{M \lambda}^{J *}(R) \mathcal{O}^{J M}(\vec{P}),
\end{equation}
where $\mathcal{D}_{M \lambda}^{J *}(R)$ is a Wigner-D matrix, and the parity $P$ is the parity of $\mathbb{O}^{J P \lambda}(\vec{P}=0)$ which depends on the parity of the gamma matrices and the number of covariant derivatives. Third, to account for the cubic symmetry of the finite volume lattice, the above-mentioned helicity operators that transform within a particular representation of $SU(2)$ need to be subduced to a particular irreducible representation (irrep) $\Lambda$ and row $\mu$ of the cubic group by,
\begin{equation}
O_{\Lambda \mu}^{[J P |\lambda|]}(\vec{P})=\sum_{\hat{\lambda}= \pm|\lambda|} \mathcal{S}_{\Lambda \mu}^{\tilde{\eta} \hat{\lambda}} \mathbb{O}^{J P \hat{\lambda}}(\vec{P}),
\end{equation}
where $\tilde{\eta} \equiv P(-1)^J$ and $\mathcal{S}_{\Lambda \mu}^{\tilde{\eta} \hat{\lambda}}$ are the subduction coefficients that can be found in Ref.~\cite{Dudek2010}.

For the construction of the $D\pi$ two-hadron operators, we propose a generalized version of the 
projection formula put forward by 
Prelovsek \textit{et. al}~\cite{Prelovsek2017},
\begin{equation}
O_{\Lambda, \mu}^{[|k|]}(\vec{P}) = \sum_{R \in G} T_{\mu, \mu}^{\Lambda}(R) R D^{(*)}_i(\vec{k}) \pi(\vec{P}-\vec{k}) R^{\dagger},
\end{equation}
where $\vec{k}$ designates the relative momentum between the two single hadrons. The spherical harmonic indices $i \in \{0\}$ for the $D$ and $i \in \{1,0,-1\}$ for the $D^*$ operators.

In this work, we consider reference frames with a center of mass three-momenta of $\frac{L}{2\pi} \vec{P} \in \{[000], [001], [011], [111], [002]\}$ and all cubic irreps with leading partial waves up to $P$-wave. Note that since $m_D \neq m_{\pi}$, $S$- and $P$-wave can mix in moving frames where parity is no longer a good quantum number.

\section{Finite volume spectra}
We perform the calculation using two clover-improved Wilson fermion ensembles generated by the CLQCD collaboration. The details of them are listed in Tab.~\ref{tab:ens}~\cite{Hu2023} and the two ensembles differ only in volume.
\begin{table}
\centering
\caption{Ensemble details}
\begin{tabular}{cccccccc}
\toprule
configuration & volume & $a$/fm & $\beta$ & $m_{\pi}$/MeV & $m_{K}$/MeV & $m_{\pi} L$ & $N_{\text{cfgs}}$ \\
\midrule
F32P30 & $32^3 \times 96$ & $0.07746(18)$ & $6.41$ & $303.2(1.3)$ & $524.6(1.8)$ & $3.8$ & $371$ \\ % beta6.41_mu-0.2295_ms-0.2050_L32x96
F48P30 & $48^3 \times 96$ & $0.07746(18)$ & $6.41$ & $303.4(9)$ & $523.6(1.4)$ & $5.7$ & $201$ \\ % beta6.41_mu-0.2295_ms-0.2050_L48x96
\bottomrule
\end{tabular}
\label{tab:ens}
\end{table}
First, the interpolating operators $\mathcal{O}_i$ constructed as described in the previous section are utilized to form the correlation matrices with definite quantum numbers
\begin{equation}
C_{i j}(t)= \sum_{t^{\prime}} \langle O_i(t+t') O_j^{\dagger}(t^{\prime}) \rangle_T
\end{equation}
where $t^{\prime}$ loops over all $96$ possible time slices to enhance the signal. Distillation method is utilized and $C_{ij}(t)$'s are estimated using relevant perambulators.

\begin{figure}[htbp]
\centering
\begin{subfigure}[b]{0.245\textwidth}
\centering
\includegraphics[width=\textwidth]{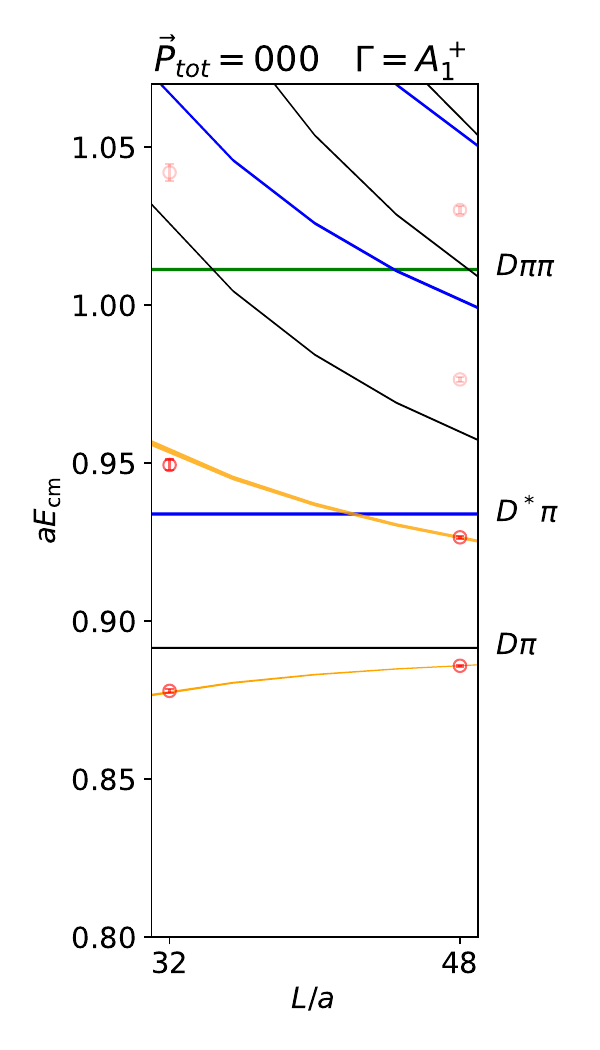}
\end{subfigure}
\hfill
\begin{subfigure}[b]{0.245\textwidth}
\centering
\includegraphics[width=\textwidth]{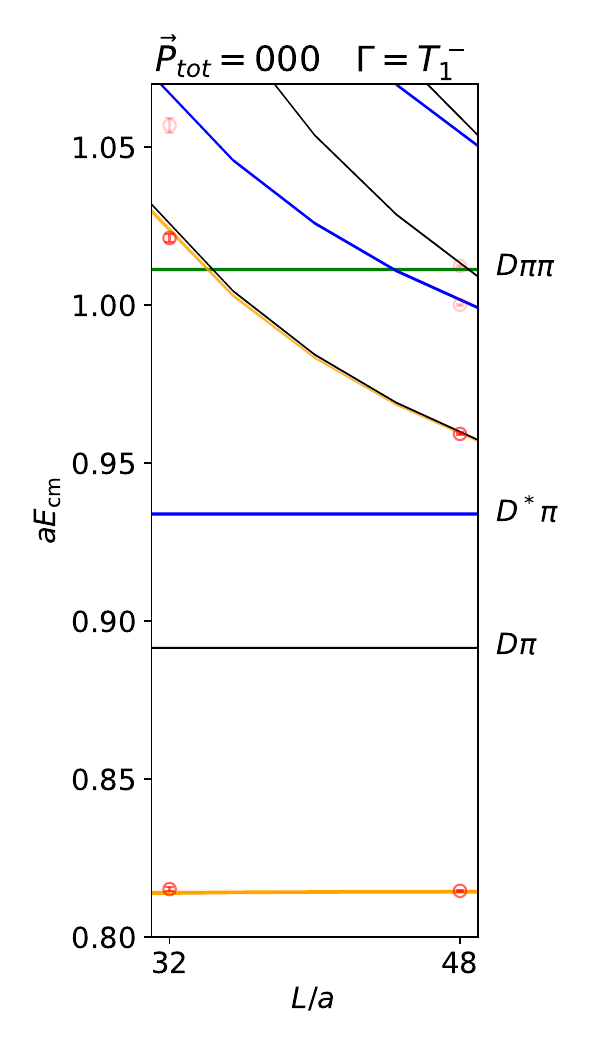}
\end{subfigure}
\hfill
\begin{subfigure}[b]{0.245\textwidth}
\centering
\includegraphics[width=\textwidth]{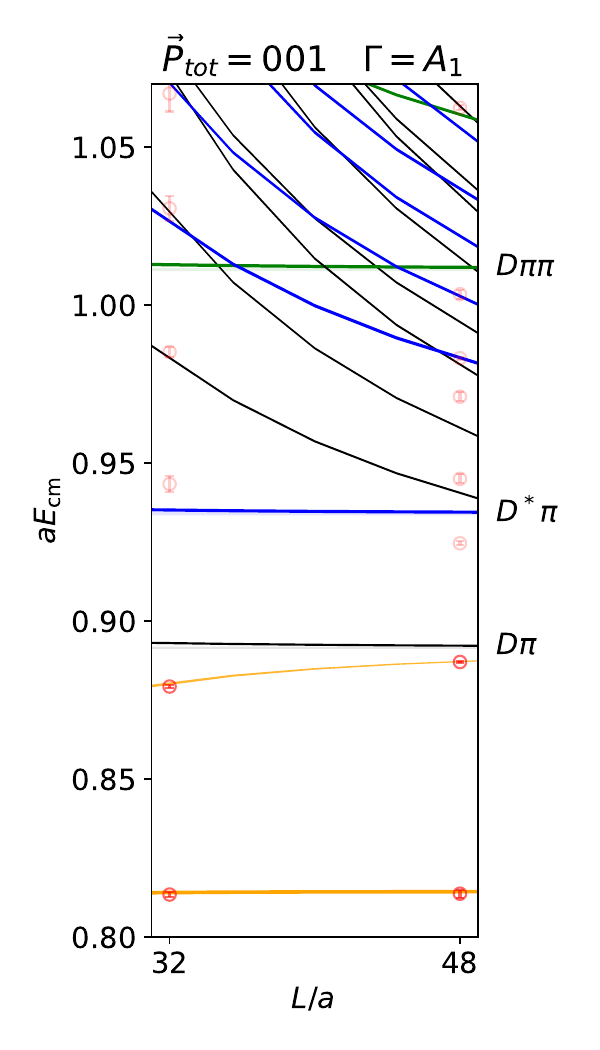}
\end{subfigure}
\hfill
\begin{subfigure}[b]{0.245\textwidth}
\centering
\includegraphics[width=\textwidth]{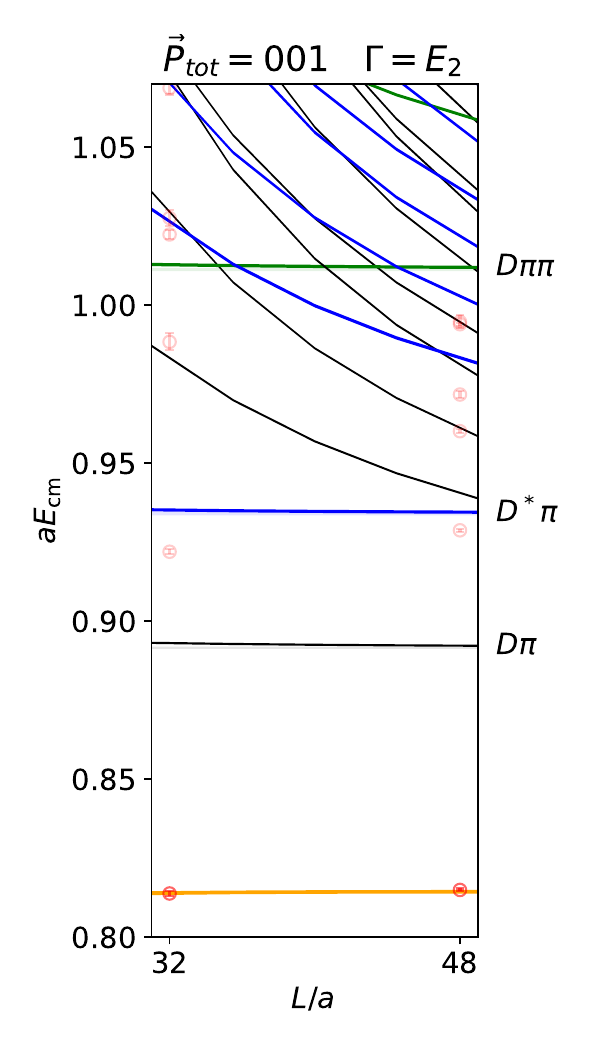}
\end{subfigure}
\caption{Examples of the finite-volume spectra for different irreps with leading partial waves being $S$- or $P$-wave. The red points are the energy levels and only the opaque points are included in the scattering analysis. The black, blue, and green curves indicate the non-interacting levels of the $D\pi$, $D^*\pi$, and the $D\pi\pi$ thresholds, respectively. The orange bands are the solutions of the L\"{u}scher's equations.}
\label{fig:spectra}
\end{figure}

In analyzing the eigenvalues of the correlation matrix using the generalized eigenvalue problem (GEVP), it is found that some eigenvalues contain non-negligible thermal pollution which can be explained by considering the decomposition of the correlation matrix. We generalize the "weighting and shifting" method as described in Ref.~\cite{Dudek2012} to the $D\pi$ system to eliminate the leading thermal state pollution. We attribute the effect to the elements whose source and sink are both $D\pi$-like 
two-body operators with the same momentum structure, which contains:
\begin{equation}
\frac{1}{Z_T} |z_{\vec{k}}^D|^2 |z_{\vec{P} - \vec{k}}^{\pi}|^2  \left[ e^{-E_\pi(\vec{P} - \vec{k}) T -(E_D(\vec{k}) - E_\pi(\vec{P} - \vec{k})) t} + e^{-E_D(\vec{k}) T -(E_\pi(\vec{P} - \vec{k}) - E_D(\vec{k})) t} \right] \\
\end{equation}
where $z_{\vec{k}}^X \equiv \langle X_{\vec{k}}^{+} | X_{\vec{k}}^{+} | \Omega \rangle$. The second thermal state is discarded since it is sub-leading, compared to the first term. The leading thermal pollution can be removed by constructing the weighted-shifted correlation matrices,
\begin{equation}
\begin{aligned}
    \tilde{C}_{ij}(t) &= e^{(E_D(\vec{k}) - E_\pi(\vec{P} - \vec{k})) t} C_{ij}(t) - e^{(E_D(\vec{k}) - E_\pi(\vec{P} - \vec{k})) (t+1)} C_{ij}(t+1).
\end{aligned}
\end{equation}
%Most elements have no such thermal state but the procedure does not affect their behavior under $t$. The spectra should be shifted up by $(E_D(\vec{k}) - E_\pi(\vec{P} - \vec{k}))$ to compensate for the weighting procedure. 
The weighted-shifted correlation matrices $\tilde{C}_{ij}(t)$ can be sent through the usual GEVP process, yielding the right energy levels. 
%Then the $n^{\text{th}}$ eigenvalues will decay like the $n^{\text{th}}$ excited state. 
By doing correlated fits of these eigenvalues, towers of energy spectra in different irreps can be obtained as shown in Fig.~\ref{fig:spectra}. 

The dispersion relation of $D$ and $\pi$ are examined and  well-described via $\sqrt{m_X^2+Z_X\vec{k}^2}$. We find $Z_\pi$ is close to $1$. However, $Z_D$ is statistically smaller than $1$, indicating a non-negligible discretization effect. High-lying levels beyond the non-interacting $D^*\pi$ thresholds are discarded in the following analysis since this requires a multi-channel description for both $D\pi$ and $D^*\pi$. In this work, only the opaque red points that are lower than the corresponding lowest non-interacting $D^*\pi$ levels are used in the L\"{u}scher's analysis.

For two-body scattering near the threshold, 
partial waves with lower $l$ dominate. It is therefore important to recognize possible $l$ quantum numbers in various irreps of the cubic group. Among the irreps studied in this work, $A_1^+$ of the center-of-mass frame and $A_1$ of all moving frame contains $S$-wave as the leading partial wave; $T_1^-$ of the center-of-mass frame, $E_2$ of $[001]$, $[111]$, $[002]$ and $B_1$, $B_2$ of $[011]$ have $P$-wave as their leading partial waves. All irreps contain $P$-wave apart from $A_1^+$ of $[000]$ as subleading partial waves. All of the irreps that contain $P$-wave have low-lying levels at around $aE_{cm} \approx 0.81$, corresponding to a vector $D^*$ bound state. For irreps having $S$-wave as the leading partial wave, we find the $D\pi$ scattering energy levels are systematically lower than their corresponding non-interacting ones, indicating a strong attractive interaction.

\section{Scattering analysis}
The finite volume spectra can be translated into the infinite-volume scattering phase shifts via L\"{u}scher's quantization condition, which can be written in the general form as~\cite{Briceno2014}.
\begin{equation}
\operatorname{det}[F^{-1}(E, \vec{P} ; L)+\mathcal{M}(E)]=0,
\end{equation}
where $F^{-1}(E, \vec{P} ; L)$ is a known function (L\"uscher's zeta function) and $\mathcal{M}(E)$ is the scattering matrix. The form of the equation varies in different moving frames and irreps. 
%To determine both the $S$- and $P$-wave at the same time,
Since L\"{u}scher's equation is an under-constrained problem, one way to proceed is to parameterize the scattering phase shifts in terms of the effective range expansion (ERE), 
\begin{equation}
k^{2l+1} \cot \delta_{l}=\frac{1}{a_{l}}+\frac{1}{2} r_{l} k^2+P_2 k^4+\mathcal{O}(k^6),
\end{equation}
where $k$ is the scattering momentum and is related to the scattering energy by
\begin{equation}
E(\vec{k})=\sqrt{m_D^2+Z_D\vec{k}^2} + \sqrt{m_\pi^2+Z_\pi\vec{k}^2}.
\end{equation}
We emphasized that the discretization effect is addressed by using the corrected dispersion relation, with $Z_X$ calculated from the single-particle energy levels.

There are $3$ energy levels below the corresponding non-interacting $D^*\pi$ levels. The scattering phase shifts are fitted with the spectra by a correlated global fit, and the fitted ERE parameters come out to be
\begin{equation}
\begin{cases}
    a_0 = 2.73(32) \ \text{fm}, \\
    r_0 = -0.564(47) \ \text{fm},
\end{cases}
\begin{cases}
    a_1 = 0.0154(56) \ \text{fm}^3, \\
    r_1 = 63(46) \ \text{fm}^{-1},
\end{cases}
\end{equation}
with $\chi^2 = 2.3$. The energy spectra for continuous volume can then be calculated from the parameterized L\"{u}scher's formula, which is also plotted as the orange bands in Fig.~\ref{fig:spectra}. The solution from the L\"{u}scher's formula describes the data at $L=32$ and $L=48$ well.

The scattering amplitudes of $S$-wave are plotted in Fig.~\ref{fig:phaseshifts1}. The energy levels used to constrain the amplitudes are shown below. We point out that in Fig.~\ref{fig:phaseshifts1:1} $\delta_0$ rises rapidly, indicating a nontrivial structure. In Fig.~\ref{fig:phaseshifts1:2} $k \cot \delta_{0}$ is plotted, where the crossing of $k \cot \delta_{0}$ and $ik$ indicates the existence of a virtual state. In Fig.~\ref{fig:phaseshifts1:3} the normalized cross section $\rho^2 |t_0|^2$ is plotted, and there is indeed an enhancement above the $D\pi$ threshold. The $P$-wave phase shift $\delta_1$ is small in the elastic zone.

\begin{figure}[htbp]
\centering
\begin{subfigure}[b]{0.32\textwidth}
\centering
\includegraphics[width=\textwidth]{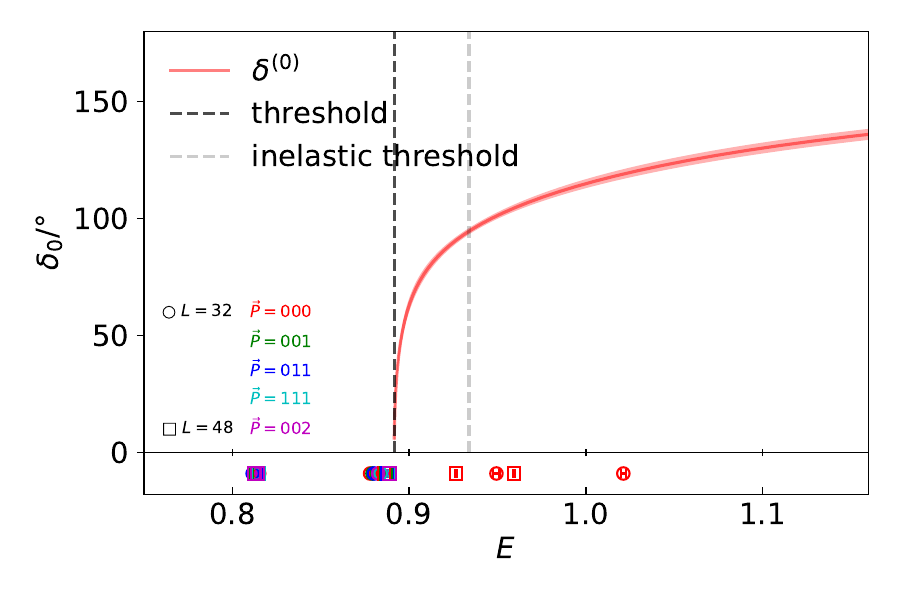}
\caption{$\delta_0$}
\label{fig:phaseshifts1:1}
\end{subfigure}
\hfill
\begin{subfigure}[b]{0.32\textwidth}
\centering
\includegraphics[width=\textwidth]{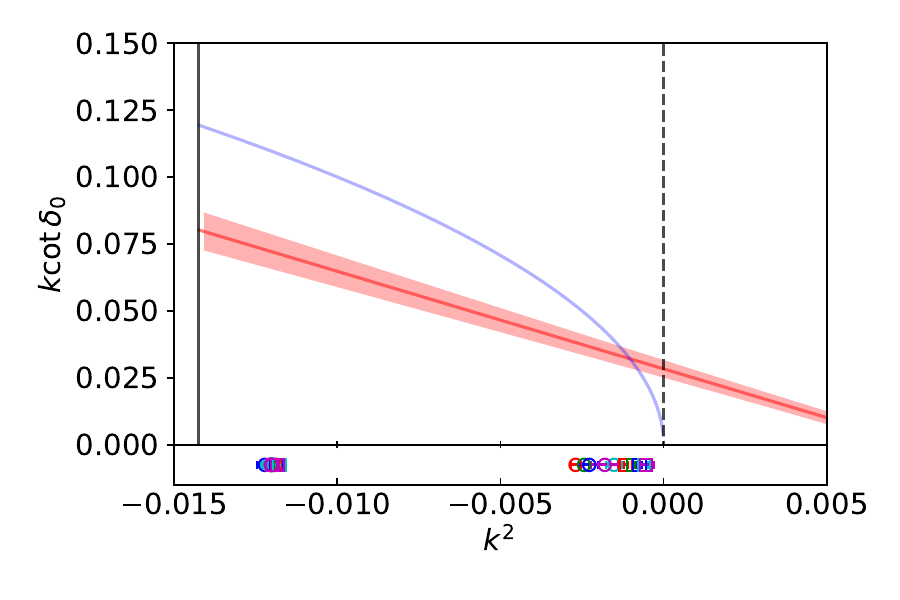}
\caption{$k \cot \delta_{0}$}
\label{fig:phaseshifts1:2}
\end{subfigure}
\hfill
\begin{subfigure}[b]{0.32\textwidth}
\centering
\includegraphics[width=\textwidth]{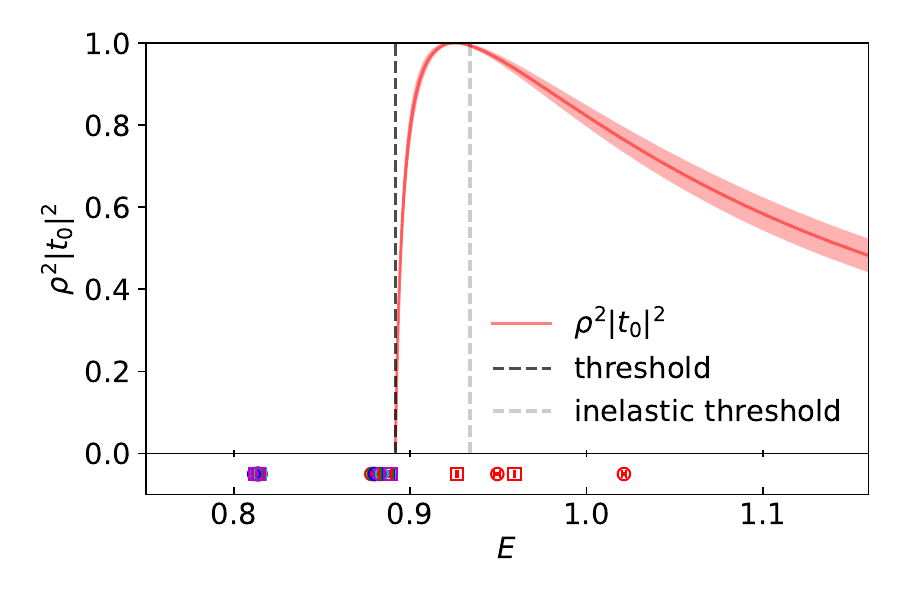}
\caption{$\rho^2 |t_0|^2$}
\label{fig:phaseshifts1:3}
\end{subfigure}
\caption{The scattering phase shifts $\delta_0$, $k \cot \delta_{0}$ and the normalized cross section $\rho^2 |t_0|^2$ of the $S$-wave. The error is purely statistical. The lower panel shows the energy levels used to constrain the amplitude.}
\label{fig:phaseshifts1}
\end{figure}

\begin{figure}[htbp]
\centering
\includegraphics[width=0.6\textwidth]{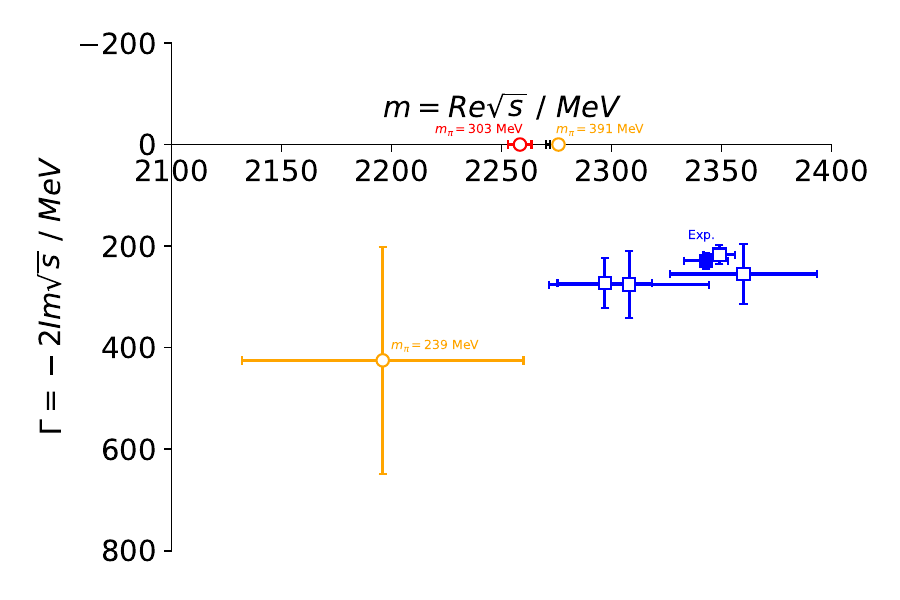}
\caption{The poles on the Riemann sheet for $S$-wave (red points). The blue points are the experimental measurements~\cite{Abe2004, Aubert2009, Aaij2015, Aaij2015a}. The solid blue point is the PDG average~\cite{Workman2022}. %The red point is the pole found in this work at $m_{\pi} = 303$ MeV. 
Yellow points are the results from the Hadron Spectrum Collaboration~\cite{Moir2016, Gayer2021} at $m_{\pi} = 391$ MeV and $m_{\pi} = 239$ MeV.}
\label{fig:pole}
\end{figure}

The scattering amplitudes are analytically continued to complex energy plane. For $S$-wave, we found a pole on the real axis below the threshold on the un-physical sheet corresponding to a virtual state and is identified to the scalar $D_0^*$ found in the experiment. The pole on the Riemann sheet is shown in Fig.~\ref{fig:pole}. Also shown are results from the Hadron Spectrum Collaboration~\cite{Moir2016, Gayer2021}, the PDG average~\cite{Workman2022} and the experimental measurements~\cite{Abe2004, Aubert2009, Aaij2015, Aaij2015a} it uses. For the $P$-wave, a pole on the real axis below the threshold on the physical sheet is found and corresponds to the vector $D^*$ bound state. A similar picture has also been observed by the Hadron Spectrum Collaboration.

Combining the results in~\cite{Moir2016, Gayer2021}, we see that as $m_{\pi}$ decreases, the $D_0^*$ pole moves from a bound state to a virtual state and finally to a resonance. The lattice results are still distant from the experiment measurements, probably due to the lattice discretization artifact and non-physical pion mass.
%inaccurate experiment analysis.

\section{Conclusions}

Using newly generated ensembles from CLQCD, the $S$- and $P$-wave $I=\frac{1}{2}$ $D\pi$ scattering phases are studied within L\"uscher's formalism. A virtual state in $S$-wave that corresponds to $D_0^*$ is identified at $m_{\pi} = 303$ MeV. Together with the results from the Hadron Spectrum Collaboration~\cite{Moir2016, Gayer2021}, we arrive at the following picture for the movement of the $D_0^*$ pole when $m_\pi$ varies: $D_0^*$ pole is a bound state at $m_{\pi} \gtrsim 391$ MeV, it then evolves into a virtual state at somewhere in $303$ MeV $\lesssim m_{\pi} \lesssim 391$ MeV, finally becomes a resonance at $m_{\pi} \gtrsim 239$ MeV. However, all three lattice results are significantly below the experimental measurement. Ongoing investigations aim to add another ensemble at $m_{\pi} \approx 210$ MeV to gain more insight into the $m_{\pi}$ dependence. The investigation with a finer lattice spacing is also planned to examine the discretization effect.

\acknowledgments
H.~Yan and C.~Liu acknowledge support from NSFC under Grant No. 12070131001, 11935017, 12293060, 12293061, 12293063. H.~Yan is grateful to L.~An, Y.~Chen, X.~Feng, B.~H\"{o}rz and J.~Wu for valuable discussions. Part of the simulations were performed on the High-performance Computing Platform of Peking University, the Southern Nuclear Science Computing Center (SNSC), and the SongShan supercomputer at Zhengzhou National Supercomputing Center.

\bibliographystyle{hb}
\bibliography{paper}

%\begin{thebibliography}{99}
%\end{thebibliography}

\end{document}